\def\t{\tilde}

\def\eV{{\rm eV}}
\def\ue3{\left| U_{e3} \right|}

\def\be{\begin{equation}}
\def\ee{\end{equation}}

\documentstyle[epsfig,12pt]{article}
\begin{document}
\baselineskip=22 pt
\setcounter{page}{1}
\thispagestyle{empty}
%\topskip  -2.  cm
\begin{flushright}
%\begin{tabular}{c c}
%&  October   3, 2005
%\end{tabular}
\end{flushright}
\vspace{0.2 cm}

\centerline{\Large\bf Embedding the Texture of the  Neutrino  Mass Matrix}
 
\centerline{\Large \bf   into  the MaVaNs Scenario} 
\vskip 0.8 cm
\centerline{{\large Mizue  Honda, $^{a}$}$\!\!\!$
\renewcommand{\thefootnote}{\fnsymbol{footnote}}
\footnote[1]{e-mail:  mizue@muse.sc.niigata-u.ac.jp}, \
{\large Ryo Takahashi$^{a}$}$\!\!\!$
\renewcommand{\thefootnote}{\fnsymbol{footnote}}
\footnote[2]{e-mail:  takahasi@muse.sc.niigata-u.ac.jp} \
{\large and  Morimitsu Tanimoto$\,^{b}$}$\!\!\!$
\renewcommand{\thefootnote}{\fnsymbol{footnote}}
\footnote[3]{e-mail: tanimoto@muse.sc.niigata-u.ac.jp}}

\vskip 0.5 cm

\centerline{$^a\!$ Graduate School of  Science and Technology,
 Niigata University,  950-2181 Niigata, Japan}

\centerline{$^b\!$ Department of Physics,
Niigata University,  950-2181 Niigata, Japan}

\vskip 2 cm
\centerline{\bf ABSTRACT}\par
\vskip 0.2 cm

We have embedded the texture of the neutrino mass matrix with  three
families into the MaVaNs scenario. We take the power-law potential of  
the acceleron field and a typical texture of  active neutrinos,
which is derived by the $D_4$ symmetry and predicts the maximal mixing of 
the atmospheric neutrino and the vanishing $U_{e3}$.  
The effect of couplings among the dark fermion and active neutrinos are 
studied by putting
the current cosmological data and the terrestrial neutrino experimental data.
It is found that the neutrino flavor mixings evolve as well as the neutrino
masses. Especially, $U_{e3}$ develops into the non-vanishing one and 
$\theta_{\rm atm}$ deviates from the maximal mixing due to  couplings 
among the dark fermion and  active neutrinos.
\newpage

\section{Introduction}

One of the most challenging questions in both cosmological and 
particle physics is the nature of the dark energy in the Universe.
At the present epoch, the energy density of the Universe is dominated 
by a dark energy component,
whose negative pressure causes the expansion of the Universe to accelerate.
 In order to clarify the origin of the dark energy, one has tried to 
understand the connection of the dark energy with particle physics.

Recently, Fardon, Nelson and  Weiner  \cite{Weiner}  proposed an idea
 of the mass varying neutrinos (MaVaNs), 
in which the neutrino couples to  the dark energy.
 Gu, Wang and Zhang \cite{Wang} also  considered the  coupling of the scalar, such as Quintessence,  to the neutrinos.  It should be also  noted  that
the variable  neutrino mass  was considered at first in 
\cite{Yanagida}, and was discussed for neutrino clouds \cite{McKellar}.

  The renewed  MaVaNs scenario \cite{Weiner}  has tried to make a
 connection between neutrinos and  the dark energy. In this  scenario,
an unknown scalar field which is called ``acceleron'' is  introduced,
and then, the neutrino mass becomes a dynamical field.
The acceleron field sits at the instantaneous minimum of its potential, 
and the cosmic expansion only modulates this minimum through changes 
in the  neutrino density.
Therefore, the neutrino mass is given by the acceleron field
and changes with the evolution of the Universe.
The cosmological parameter $w$ and
 the dark energy also evolve with  the neutrino mass.
Those evolutions depend on a model of the scalar potential strongly. Typical 
examples of the potential have been discussed by Peccei \cite{Peccei}. 

The  MaVaNs scenario leads to interesting  phenomenological results.
 The neutrino oscillations may be a probe of the dark energy 
\cite{Kaplan,Mini}.
 The baryogenesis \cite{Wang,Bi,Gu}, the cosmo MSW effect of  neutrinos 
\cite{Hung} and the solar neutrino \cite{SUN1,SUN2} 
have been studied in the context of this  scenario.
 Cosmological  discussions of  the  scenario are also presented 
\cite{Zhang,Right,stability,WZ,Supernova}.
The extension to the supersymmetry have been presented in
 ref. \cite{our,SUSY}.

Now one needs to construct the realistic MaVaNs scenario with three families,
which is consistent with  the terrestrial neutrino experimental data 
\cite{solar,Kamland,atmospheric,chooz,lisi}.
In order to get a realistic  MaVaNs scenario,
we have  embedded a texture of the neutrino mass matrix with three families
 into  the MaVaNs scenario
 and examined neutrino masses and flavor mixings \cite{mns}.
   We take a typical  texture, which is derived
 by $D_4$ symmetry and predicts the  maximal  flavor  mixing  
  and the vanishing one in the lepton sector.
The effect of couplings among the dark fermion 
and active neutrinos are studied by putting  the current cosmological data 
\cite{Data} and the terrestrial neutrino experimental data 
\cite{solar,Kamland,atmospheric,chooz,lisi}.
It is found that neutrino flavor mixings evolve as well as neutrino masses.
Especially,  $U_{e3}$ develops into the non-vanishing one
 and $\theta_{\rm atm}$ deviates from the maximal mixing
due to the couplings among the dark fermion and active neutrinos.

In section 2, we present the formulation of the MaVaNs  scenario with 
three families, and in section 3, we study evolutions
of neutrino masses and flavor mixings. The section 4 devotes to the summary.

%%%%%%%%%%%%%%%%%%%%%%%%%%%%%%%%%%%%%%%%%%%%%%%%%%%
%%%%%%%%%%%%%%%%%%%%%%%%%%%%%%%%%%%%%%%%%%%%%%%%%%%
\section{Dark Energy and Three Active Neutrino Masses}

In the MaVaNs scenario,   one considers a dark energy sector 
consisting of an acceleron field, $\phi_a$ and 
a dark fermion $n$. Only left-handed neutrinos are supposed to  couple
to  the dark sector. There are two constraints on the scalar
potential of  the  acceleron field $\phi_a$. 
The first one comes from the observation of the Universe, 
which  is that the present dark energy density is about $0.7\rho_c$, 
$\rho_c$ being  the  critical density.
 Since the dark energy is assumed to be the sum of the energy densities 
 of  neutrinos  and the scalar potential in the MaVaNs scenario,
\begin{equation}
 \rho_{\mbox{dark}}=\rho_\nu +V(\phi_a) \ ,
\end{equation} 
\noindent the first constraint turns to 
\begin{equation}
\rho^0_\nu+V(\phi _a^0)=0.7\rho _c \ ,
\label{con1}
\end{equation}
where ``0'' represents a value at the present epoch,
 and  $70\%$ is taken for the dark energy in the Universe. 

The second one
comes from the fundamental assumption in this scenario, which  is that
$\rho _{\mbox{dark}}$ is stationary with respect to variations in the
neutrino mass. This assumption is represented by
\begin{equation}
\frac{\partial\rho _\nu}{\partial \sum m_{\nu i}}+
\frac{\partial V(\phi_a(m_{\nu i} ))}{\partial  \sum m_{\nu i}}=0 \ .
\label{stationary1}
\end{equation}
For our purpose it suffices to consider the neutrino mass as a
function of the cosmic temperature $T$ \cite{Peccei}.
Since one can write down generally
\begin{equation}
\rho_\nu=T^4 \sum_{i=1}^3 F(\xi_i) \ ,  \qquad \xi_i=\frac{m_{\nu i}}{T} \ ,
\end{equation}
%%%%%%%%%%%%%%% the stationary condition %%%%%%%%%%%%%%%%%%%%%%%%%%%
  the stationary condition eq. (\ref{stationary1}) turns to 
\begin{equation}
T^4 \sum_{i=1}^3\frac{\partial F}{\partial\xi_i}\frac{\partial \xi_i}{\partial \phi_a}+
\frac{\partial V(\phi_a)}{\partial \phi_a}=0 \ ,
\label{stationary2}
\end{equation}
where  
\begin{equation}
F(\xi_i )=\frac{1}{\pi ^2}\int _0^\infty \frac{dyy^2\sqrt{y^2+\xi_i
^2}}{e^y+1}\ .
\label{Func}
\end{equation}
%%%%%%%%%%%%%%%%%%%%%%%%%%%%%%%%%%%%%%%%%%%%%%%%%%%%%%%%%%%%%%%%%%%%5
We can obtain the time evolution of  neutrino masses from the relation of
 eq.(\ref{stationary2}). 

Since the stationary condition should be also satisfied at the present epoch,
 the second constraint on the scalar potential is
%%%%%%%%%%%%%%% Two constraints %%%%%%%%%%%%%%%%%%%%%%%%%%%%%%%%%%%%%%
\begin{eqnarray}
\left.\frac{\partial V(\phi _a)}{\partial \phi_a}\right|_{\phi_a=\phi_a^0}
=-T^4 \sum_{i=1}^3 \left .
\frac{\partial F(\xi_i)}{\partial\xi_i}\frac{\partial \xi_i}{\partial\phi_a}
\right |_{m_{\nu i}(\phi_a)=m^0_{{\nu i}}(\phi^0_a), \ T=T_0} \ ,
\label{con2}
\end{eqnarray}
where $T_0=1.69\times 10^{-4} \eV$.
%%%%%%%%%%%%%%%%%%%%%%%%%%%%%%%%%%%%%%%%%%%%%%%%%%%%%%%%%%%%%%%%%%%%%%
It is found that the gradient of the scalar potential should be 
negative and very small.

%%%%%%%%%%%%%%%%%%%%%%%%%%%%%%%%%%%%%%%%%%%%%%
%%%%%%%%%%%%%%%  Equation of State %%%%%%%%%%%
%%%%%%%%%%%%%%%%%%%%%%%%%%%%%%%%%%%%%%%%%%%%%%
One can also calculate
the  equation of state parameter $w$ as follows:
\begin{equation}
 w+1=\frac{4- h(T)}{3\left[ 1+
\frac{V(\phi_a)}{T^4 \sum F(\xi_i)}\right ]} \ ,
\label{w}
 \end{equation}
\noindent  where
\begin{equation}
 h(T)=\frac{\sum \xi_i\frac{\partial F(\xi_i)}{\partial \xi_i}}
  { \sum F(\xi_i)} \ .
 \end{equation}

In order to calculate the evolution of neutrino masses and $w$,
we assume the power-low potential for the scalar potential $V(\phi_a)$
 as follows:
\begin{equation}
 V(\phi_a)=  A \left ( \frac{\phi_a}{\phi_a^0}\right )^k \ ,
\label{V0}
 \end{equation}
\noindent where $A$ and $k$ are fixed by the condition of
eqs.(\ref{con1}) and (\ref{con2})
if  the  magnitude of the acceleron field at present , $\phi_a^0$ is given.
Then, we can  calculate  evolutions of neutrino masses and $w$.

%%%%%%%%%%%%%%%%%%%%%%%%%%%%%%%%%%%%%%%%
%%%%%%%%%%%%%%%%%%%%%%%%%%%%%%%%%%%%%%%%
\section{Neutrinos Masses  and Flavor Mixings}

 Let us discuss the neutrino mass matrix.
 Suppose  only  left-handed neutrinos  couple to the dark sector, 
which consists of  an acceleron field, $\phi_a$ and  a dark fermion $n$.
 A Lagrangian for the dark  sector and the neutrino sector is given as

\begin{equation}
{\mathcal L}={\overline\nu_{L\alpha}}m^\alpha_D n+\lambda \phi_a nn +
 {\overline\nu_{L\alpha}} M^{\alpha\beta}_{D}\nu_{R\beta} 
+\nu_{R\alpha}^T M^{\alpha\beta}_{R} C^{-1}\nu_{R\beta}+ h.c.\ ,
\end{equation}
\noindent
 $\nu_L$ and $\nu_R$ are the left-handed and the right-handed neutrinos,
respectively. Since we consider three families of  active neutrinos,
  mass matrices $m^\alpha_D$, $M^{\alpha\beta}_{D}$ and
 $M^{\alpha\beta}_{R}$ are $3\times 1$,  $3\times 3$ and  $3\times 3$
matrices, respectively.
Then, the neutrino mass matrix is given as the $7\times 7$ matrix
\begin{equation}
 M=   \left(
    \begin{array}{ccc}
     0    & m_D & M_D \\ m^T_D & \lambda \phi_a & 0 \\ M_D^T  & 0 & M_R
    \end{array}
   \right) \  , 
\label{mass77}
 \end{equation}
\noindent
 in the $(\nu_L, n , \nu_R)$ basis.
We   assume the right-handed Majorana mass scale $M_R$ to be much higher 
than the Dirac neutrino mass scale $M_D$, and  
$\lambda \phi_a$ to be  much  higher than  the scale of $m_D$.  Then,
 the effective neutrino mass matrix is  approximately given as
\begin{equation}
 M_\nu =  M_D M_R^{-1} M_D^T + \frac{m_D m_D^T}{\lambda \phi_a} \ .
\label{effectivemass}
\end{equation}
\noindent
Furthermore, we have one sterile neutrino with the mass $\lambda \phi_a$
and heavy three right-handed Majorana neutrinos.

The first term in the right hand side of eq.(\ref{effectivemass})
is the ordinary neutrino seesaw mass matrix, which is denoted by  $\t M_\nu$,
 and so it depends on the flavor model of neutrinos.   On the other hand,
the second term originates from couplings of the left-handed neutrinos
 to  the dark  sector.
The matrix $m_D$ is parametrized as 
\begin{equation}
 m_D=   D \left(
    \begin{array}{ccc}
     a  \\ b \\ c
    \end{array}
   \right) \  , 
\label{coupling}
 \end{equation}
\noindent 
 where  coefficients $a$, $b$ and $c$ are introduced
 to indicate relative couplings  to  three flavors.
%%%%%%%%%%%%%%%%%%%%%%%%%%%%%%%%%%%%%%%%%%%
Then, the second term  is written  as
\begin{equation}
  \frac{m_D m_D^T}{\lambda \phi_a}=  \frac{D^2}{\lambda \phi_a} \left(
    \begin{array}{ccc}
      a^2   & a b & a c\\  a b& b^2 & b c \\ a c  & b c & c^2
    \end{array}
   \right) \  .
 \end{equation}
\noindent
% where $a$, $b$ and $c$ are order one coefficients.
 In this paper, we assume that  the ordinary neutrino seesaw mass matrix
$\t M_\nu$ dominates the effective neutrino mass matrix $M_\nu$. Therefore,
 the contribution from the dark sector is the next-leading term.
 The first  term depends on the model of the neutrino mass matrix.
In order to find the effect of  the  acceleron on  neutrino masses 
and mixings, we take the  following typical texture of  neutrinos
 in the flavor basis of the charged lepton:
\begin{equation}
\t M_\nu = \left( \matrix{
 X &  Y &  Y \cr  Y &  W &  U\cr  Y &  U &  W \cr}
\right) \ ,
\label{D4}
\end{equation}
\noindent
which leads to the neutrino flavor mixing matrix:
\begin{equation}
\t U = \left( \matrix{
 c_{12} &  s_{12} &  0 \cr  
-\frac{1}{\sqrt{2}}s_{12} &  \frac{1}{\sqrt{2}} c_{12} & 
 -\frac{1}{\sqrt{2}} \cr 
 -\frac{1}{\sqrt{2}} s_{12} &  \frac{1}{\sqrt{2}} c_{12}  &  
 \frac{1}{\sqrt{2}}\cr}
\right) \ ,
\label{D4mix}
\end{equation}
\noindent with arbitrary three neutrino masses $\t m_1$, $\t m_2$ 
and $\t m_3$.  Here, $c_{12}$ and $s_{12}$ denote $\cos\theta_{12}$
and  $\sin\theta_{12}$, respectively.
 The mass matrix of eq.(\ref{D4}) was derived by the discrete symmetry
$D_4$ and examined phenomenologically \cite{Grimus}. 
 This mass matrix gives  $U_{e3}=0$ and $U_{\mu 3}=-1/\sqrt{2}$ as seen 
in eq.(\ref{D4mix}). On the other hands, $\theta_{12}$ is an arbitrary
mixing angle. Therefore, this neutrino mass matrix is attractive one
because it  guarantees the one maximal flavor mixing and one
 vanishing  flavor mixing.
 % The breaking of the $D_4$ flavor symmetry should be taken into
 %in order to get the non-vanishing  $U_{e3}$ and the deviation
 %from the maximal flavor mixing.

In this paper, we discuss the effect of couplings of the left-handed neutrinos to the  dark fermion on the neutrino mass matrix with the flavor symmetry.
Since the contribution of the dark fermion depends on 
couplings  $a,\ b,\ c$, which denote  flavor couplings, we examine 
%%%%%%%%%%%%%%%%%%%%%%%%%%%%%%%%%%%%%%%%%%%%%%%%%%%%%%%%%%%%%%%
  typical three cases of $a=b=c=1$ (the flavor blind),  
 $a=1, b=c=0$ (the first family coupling) and 
  $a=b=0, c=1$ (the third family coupling).
The other choices of $a,b,c$ 
 do not provide drastic changes  of reslts in these three cases.

%%%%%%%%%%%%%%%%%%%%%%%%%%%%%%%%%%%%%%%%%%%%%%%%%%%%%%%%%%%%%%%
\subsection{The case of $a=b=c=1$  (the flavor blind)}
At first, we consider the case that  couplings among the left-handed 
neutrinos and the dark fermion are  the flavor  blind.
Therefore, we take $a=b=c=1$ in eq.(\ref{coupling}), which leads to the
effective neutrino mass matrix 
\begin{equation}
 M_\nu=    \left( \matrix{
 X &  Y &  Y \cr  Y &  W &  U\cr  Y &  U &  W\cr}
\right) + \frac{D^2}{\lambda \phi_a} \left( \matrix{
 1 &  1 &  1 \cr  1 &  1 &  1\cr  1 &  1 &  1\cr}
\right)    \ . 
\label{effectiveabc}
\end{equation}
\noindent
 Since the structure of the neutrino mass matrix is not changed 
 by the dark sector as seen in  eq.(\ref{effectiveabc}),
the flavor mixings do not deviate from $\t U_{e3}=0$ and 
$\t U_{\mu 3}=-1/\sqrt{2}$.
 Assuming that the contribution of the dark sector in eq.(\ref{effectiveabc})
is small  compared with the active neutrino mass matrix, 
we get the mass eigenvalues $m_i(i=1\sim 3)$ and eigenvectors  $u_i$
in the first order perturbative expansion as follows:
\begin{eqnarray}
m_i &=& \t m_i+  M_{ii}^{'(1)} \ , \nonumber \\
u_i &=& \t u_i + u_i^{(1)} \ ,
\label{sol}
\end{eqnarray}
\noindent
where
\begin{eqnarray}
M_{ij}^{'(1)}&=&\frac{D^2}{\lambda \phi_a}\  
 \left( \matrix{
 \t U_{ei}^{*} &  \t U_{\mu i}^{*} &  \t U_{\tau i}^{*} \cr}\right)
  \left( \matrix{
 1 &  1 &  1 \cr  1 &  1 &  1\cr  1 &  1 &  1\cr}
\right)  
 \left( \matrix{
  \t U_{ej}\cr  \t U_{\mu j} \cr  \t U_{\tau j} \cr}
\right)  
\nonumber\\
&=&\frac{D^2}{\lambda \phi_a}\  
(\t U_{ei}^{*}+ \t U_{\mu i}^{*}+\t U_{\tau i}^{*})\ 
 (\t U_{ej}+ \t U_{\mu j}+\t U_{\tau j}) \ , \nonumber \\
 u_i^{(1)} &=& C_{i1} \t u_1+ C_{i2} \t u_2+ C_{i3}\t u_3\ ,
\end{eqnarray}
\noindent
with 
\begin{eqnarray}
C_{ij}=\frac{M_{ij}^{'(1)}}{\t m_i-\t m_j}\ , \qquad (i \not= j) \ ,
\qquad\qquad C_{ii}=0 \ ,
\end{eqnarray}
\noindent
and 
\begin{eqnarray}
\t u_1= \left (\matrix{c_{12}\cr -\frac{s_{12}}{\sqrt{2}}\cr
   -\frac{s_{12}}{\sqrt{2}}} \right ) \ ,
\qquad 
\t u_2= \left (\matrix{s_{12}\cr \frac{c_{12}}{\sqrt{2}}\cr
   \frac{c_{12}}{\sqrt{2}}} \right ) \ ,
\qquad 
\t u_3=\left (\matrix{0\cr -\frac{1}{\sqrt{2}}\cr\frac{1}{\sqrt{2}}}
\right ) \ .
\end{eqnarray}
\noindent
Since we get 
 \begin{eqnarray}
M^{'(1)}&=&\frac{D^2}{\lambda \phi_a}\ 
\left( \matrix{
(c_{12}-\sqrt{2}s_{12})^2&(c_{12}-\sqrt{2}s_{12})(\sqrt{2}c_{12}+s_{12})& 0\cr
(c_{12}-\sqrt{2}s_{12})(\sqrt{2}c_{12}+s_{12})&(s_{12}+\sqrt{2}c_{12})^2&0\cr
 0 & 0&0\cr}
\right) \ , 
\end{eqnarray}
mass eigenvalues and  flavor mixings  are written 
in the first order approximation:
\begin{eqnarray}
m_1 &=& \t m_1+ \frac{D^2}{\lambda \phi_a} (c_{12}-\sqrt{2}s_{12})^2  \ ,
        \nonumber \\
m_2 &=& \t m_2+ \frac{D^2}{\lambda \phi_a} (s_{12}+\sqrt{2}c_{12})^2  \ ,
        \nonumber \\
m_3 &=& \t m_3\ ,   \label{solabc}  \\
U_{e2}&=&s_{12}+ \frac{D^2}{\lambda \phi_a}
\ c_{12}\ \frac{(c_{12}-\sqrt{2}s_{12})(\sqrt{2}c_{12}+s_{12})}
{ \t m_2- \t m_1} \ , \nonumber\\
U_{\mu 3}&=& -\frac{1}{\sqrt{2}}\ , \nonumber\\
U_{e3}&=&0\ , \nonumber
\end{eqnarray}
\noindent
 where $U_{e3}$ and $U_{\mu 3}$ are free  
from the contribution  of  the dark sector.

%%%%%%%%%%%%%%%%%%%%%%%%%%%%%%%%%%%%%%%%%%%5
%%%%%%%%%%%%%%%%%%%%%%%%%%%%%%%%%%%%%%%%%

 Since the parameters $\t m_i (i=1\sim 3)$ and $s_{12}$ are arbitrary,
 these  are taken to be consistent with the experimental data
  in the $90\%$ CL limit \cite{solar,Kamland,atmospheric,lisi}:
\begin{eqnarray}
 0.33 \le  \tan^2 \theta_{\rm sun} \le 0.49  \ , \ \
&&7.7\times 10^{-5} \le  \Delta m_{\rm sun}^2 \le
8.8\times 10^{-5}~\rm{eV}^2 \ , \nonumber \\
0.92 \le \sin^2 2\theta_{\rm atm}  \ , \ \
&& 1.5\times 10^{-3} \le  \Delta m_{\rm atm}^2 \le 3.4 \times 10^{-3}~
 \rm{eV}^2  \ .
\label{data}
\end{eqnarray} 
The bound obtained by the reactor neutrinos \cite{chooz} 
$\theta_{\rm reactor}\leq 12^\circ$ is also taken in our study.

Assuming the normal hierarchy of neutrino masses, we take the following
typical neutrino masses at present epoch:
%%%%%%%%%%%%%%%%%%%%%%%%%%%%%%%%%%%%%%%%%%%%%%%%%%
\begin{eqnarray}
m_3=0.05\ {\rm eV}\ , \qquad  m_2=0.01\ {\rm eV}\ ,
\qquad m_1=0.0045\ {\rm eV} \ . 
\label{typical}
\end{eqnarray}
%%%%%%%%%%%%%%%%%%%%%%%%%%%%%%%%%%%%%%%%%%%%%%%%%%
\noindent These values lead to $\Delta m_{\rm atm}^2=2.4\times 10^{-3}\ \eV$ 
and  $\Delta m_{\rm sun}^2=8.0\times 10^{-5}\ \eV$,
which are consistent with the experimental values in eq.(\ref{data})
%%%%%%%%%%%%%%%%%%%%%%
\footnote{There are other choices for $m_2$ and $m_1$. For example,
the set of $m_2=0.009\ {\rm eV}$ and $m_1=0.001\ {\rm eV}$ also 
gives  $\Delta m_{\rm sun}^2=8.0\times 10^{-5}\ \eV$. However, 
the numerical results in Table 1  do  not so change.}.
%%%%%%%%%%%%%%%%%%%%%
On the other hand, 
 since we expect that $U_{e2}$ does not  deviate from $s_{12}$ so much,
 we take a typical value $s_{12}=0.5$, which gives the  relevant value of 
$U_{e2}$.
 By using these values, we examine the effect of the dark sector
on  neutrino masses and  flavor mixings.
Evolutions of masses and mixings depend on the parameter of the 
dark sector   $D^2/\lambda \phi^0_a$, which is unknown one.
  We show the numerical results for three cases of
%%%%%%%%%%%%%%%%%%%%%%%%%%%%%%%%%%%%%%%%%%%%%%%%%%
\begin{eqnarray}
\frac{D^2}{\lambda \phi^0_a}= 10^{-3}\ \eV \ , \quad 5\times 10^{-4}\ \eV \ ,
 \quad  10^{-4} \ {\rm eV} \ ,
\label{dark}
\end{eqnarray}
because the effect of the dark sector is tiny  in the case of  
$D^2/\lambda \phi^0_a\leq 10^{-4}\eV$. On the other hand,
the speed of sound becomes imaginary in the case of  
$D^2/\lambda \phi^0_a\geq 10^{-3}\eV$, which will be discussed
 in the section 4.

%%%%%%%%%%%%%%%%%%%%%%%%%%%%%%%%%%%%%%%%%%%%%%%%%%
For each case of eq.(\ref{dark}), we solve  equations in eq.(\ref{solabc})
 at present epoch, and then  we can get $\t m_1$,  $\t m_2$,  $\t m_3$,
 $U_{e2}$,  $U_{\mu 3}$ and  $U_{e3}$.
The numerical results are presented in Table 1.
%%%%%%%%%%%%%%%%%%%%%%%%%%%%%%%%%%%%%%%%%%%%%%%%%%
It is noticed  that  $m_3$,  $U_{\mu 3}$ and $U_{e3}$ are free from 
the effect of the dark sector, while $m_2$ and $U_{e2}$ have 
the seizable contribution from the dark sector,  and $m_1$ has the tiny one.  
Therefore,  $\Delta m_{\rm atm}^2$, $\Delta m_{\rm sun}^2$ and $U_{e2}$ 
are time dependent in the universe.

On the other hand,  the sterile neutrino mass  is predicted  
for each case of eq.(\ref{dark}) as  
$1.0 D^2 {\rm keV}$, $2.0 D^2 {\rm keV}$ and  $10 D^2 {\rm keV}$, 
respectively, 
which depends on the unknown parameter $D$  in the $\rm eV$ unit.

%%%%%%%%%%%%%%%%%%%%%%%%%%%%%%%%%%%%%%%%
Since we know  $\phi_a$ dependence of  neutrino masses,
we get the parameter $A$ and $k$ in the scalar potential $V(\phi_a)$
of eq.(\ref{V0}).
The conditions of eqs. (\ref{con1}) and (\ref{con2}) turn to be
\begin{eqnarray}
A&=&0.7\rho_c - \rho^0_\nu
=2.79\times 10^{-11}{\rm eV} \ , \nonumber\\
k&=&T^3 \sum_{i=1}^3 \left .
\frac{\partial F(\xi_i)}{\partial\xi_i}
\right |_{m_{\nu i}(\phi_a)=m^0_{\nu i}(\phi^0_a), \ T=T^0}
\times  \frac{D^2}{A\lambda \phi_a^0} [(c_{12}-\sqrt{2}s_{12})^2+
(s_{12}+\sqrt{2}c_{12})^2] \nonumber\\
&=&3n_\nu^0  \frac{D^2}{A\lambda \phi_a^0} \ ,
\label{Ak1}
\end{eqnarray}
where $n_{\nu}^0=8.82\times 10^{-13}{\rm eV^3}$ is taken 
for  the neutrino number density at present epoch .
The values of $k$ are also  summarized in Table 1.
%%%%%%%%%%%%%%%%%%%%%%%%%%%%%%%%%%%%%%%%%%%%%%%%%%%%%%%%%%%%%
\begin{table}[htb]
\begin{center}
\begin{tabular}{|r|r|r|r|}
\hline
 & & &  \\
Couplings$\quad$ & $\frac{D^2}{\lambda \phi_a}\ (\eV)$ & $\t m_3(\eV)$ \quad
  $\t m_2(\eV)$ \ \  $\t m_1 (\eV)\qquad$ \   $k\qquad$ &
             $U_{e2}$\quad\ \   $U_{\mu 2}$\quad\ \ $U_{e3}\qquad $  \\ 
 & & &  \\
\hline
 & & &  \\
 $a=b=c=1$
 & $10^{-3}$ &
 $0.05\ \  $\quad $0.0070\ \  $\quad $0.0045\quad$ $9.47\times 10^{-5}$ & 
 $0.59$\quad $-1/\sqrt{2}$ \quad $0\qquad\ $\\ 
   & $5\times 10^{-4}$ &
 $0.05\ \  $\quad $0.0085\ \  $\quad $0.0045\quad$ $4.73\times 10^{-5}$ & 
 $0.53$\quad $-1/\sqrt{2}$ \quad $0\qquad\ $\\ 
 & $10^{-4}$ &
 $0.05\ \  $\quad $0.0097\ \  $\quad $0.0045\quad$ $9.47\times 10^{-6}$ & 
 $0.51$\quad $-1/\sqrt{2}$ \quad $0\qquad\  $\\ 
 & & &  \\
\hline
 & & &  \\
 $a=1\qquad\quad\ $
 & $10^{-3}$ &
 $0.05\ \  $\quad $0.0098\ \  $\quad $0.0037\quad$ $3.16\times 10^{-5}$ & 
 $0.56$\quad $-1/\sqrt{2}$ \quad $0\qquad\ $\\ 
  $b=c=0\quad\ \ $  & $5\times 10^{-4}$ &
 $0.05\ \  $\quad $0.0099\ \  $\quad $0.0041\quad$ $1.58\times 10^{-5}$ & 
 $0.53$\quad $-1/\sqrt{2}$ \quad $0\qquad\  $\\ 
 & $10^{-4}$ &
 $0.05\ \  $\quad $0.0010\ \  $\quad $0.0044\quad$ $3.16\times 10^{-6}$ & 
 $0.51$\quad $-1/\sqrt{2}$ \quad $0\qquad \ $\\ 
 & & &  \\
\hline
 & & &  \\
 $a=b=0\quad\  $
 & $10^{-3}$ &
 $0.0495\ \  $\quad $0.0096\ \  $\quad $0.0044\quad$ $3.16\times 10^{-5}$ & 
 $0.46$\quad $-0.698$ \quad $0.0006$\\ 
  $c=1\qquad\quad \  $  & $5\times 10^{-4}$ &
 $0.0498\ \  $\quad $0.0098\ \  $\quad $0.0044\quad$ $1.58\times 10^{-5}$ & 
 $0.48$\quad $-0.703$ \quad $0.0003$\\ 
 & $10^{-4}$ &
 $0.0499\ \  $\quad $0.0099\ \  $\quad $0.0045\quad$ $3.16\times 10^{-6}$ & 
 $0.49$\quad $-0.706$ \quad $0.0001$\\ 
 & & &  \\
\hline
\end{tabular}
\end{center}
\caption{The summary of output parameters and predictions of mixings.
The input parameters are taken
as $m_3=0.05\ \eV$, $m_2=0.01\ \eV$, $m_1=0.0045\ \eV$ and $s_{12}=0.5$}
\end{table}
%%%%%%%%%%%%%%%%%%%%%%%%%%%%%%%%%%%%%%%%%%%%
%%%%%%%%%%%%%% a=1, b=c=0 %%%%%%%%%%%%%%%%%%
%%%%%%%%%%%%%%%%%%%%%%%%%%%%%%%%%%%%%%%%%%%%
\subsection{The case of $a=1, b=c=0$ (the first family coupling)}
Let us consider the case that flavor dependent couplings 
of the left-handed  neutrinos to the dark fermion.
The case of  $a=1,\ b=c=0$ in eq.(\ref{coupling}) leads to the
effective neutrino mass matrix 
%%%%%%%%%%%%%%%%%%%%%%%%%%%%%%%%%%%%%%%%%%%%
 \begin{eqnarray}
M_{ij}^{'(1)}&=&\frac{D^2}{\lambda \phi_a}\ 
\left( \matrix{
c_{12}^2& c_{12} s_{12}& 0\cr
c_{12}s_{12}& s_{12}^2&0\cr
 0 & 0&0\cr}
\right) \ , 
\end{eqnarray}
\noindent which gives  mass eigenvalues and  flavor mixings 
in the first order approximation:
\begin{eqnarray}
m_1 &=& \t m_1+ \frac{D^2}{\lambda \phi_a} c_{12}^2  \ ,
        \nonumber \\
m_2 &=&\t m_2+ \frac{D^2}{\lambda \phi_a} s_{12}^2  \ ,
        \nonumber \\
m_3 &=& \t m_3\ ,   \label{sola}  \\
U_{e2}&=&s_{12}+ \frac{D^2}{\lambda \phi_a}
\  \frac{c_{12}^2 s_{12}}{\t m_2- \t m_1} \ , \nonumber\\
U_{\mu 3}&=& -\frac{1}{\sqrt{2}}\ , \nonumber\\
U_{e3}&=&0\ . \nonumber
\end{eqnarray}
\noindent Taking same input parameters  in the case of $a=b=c=1$,
 we get numerical results as shown in Table 1.
%%%%%%%%%%%%%%%%%%%%%%%%%%%%%%%%%%%%%%%%%%%%%%%%%%
In this case, 
 $m_3$,  $U_{\mu 3}$ and $U_{e3}$ are free from the effect of 
the dark sector as well as the case of $a=b=c=1$. 
On the other hand,   $m_1$, $m_2$ and  $U_{e2}$ 
have the seizable contribution from the dark sector. Furthermore, we get 
\begin{eqnarray}
A=2.79\times 10^{-11}{\rm eV} \ , 
\label{Ak2}
\end{eqnarray}
\noindent which is the same one in eq.(\ref{Ak1}), 
and $k$ is summarized in Table 1.

%%%%%%%%%%%%%%%%%%%%%%%%%%%%%%%%%%%%%%%%%%%%
%%%%%%%%%%%%%% a=b=0, c=1 %%%%%%%%%%%%%%%%%%
%%%%%%%%%%%%%%%%%%%%%%%%%%%%%%%%%%%%%%%%%%%%
 \subsection{The case of $a=b=0, c=1$ (the third family coupling)}
We discuss another case of couplings.
Since the case of  $a=b=0,\ c=1$ in eq.(\ref{coupling}) 
leads to the effective neutrino mass matrix 
 \begin{eqnarray}
M_{ij}^{'(1)}&=&\frac{1}{2}\frac{D^2}{\lambda \phi_a}\ 
\left( \matrix{
s_{12}^2& -c_{12}s_{12}& -s_{12}\cr
 -c_{12}s_{12}&c_{12}^2& c_{12}\cr
 -s_{12} & c_{12}& 1\cr}
\right) \ , 
\end{eqnarray}
we get mass eigenvalues and flavor mixings in the first order approximation:
\begin{eqnarray}
m_1 &=& \t m_1+ \frac{1}{2}\frac{D^2}{\lambda \phi_a} s_{12}^2  \ ,
        \nonumber \\
m_2 &=& \t m_2+ \frac{1}{2} \frac{D^2}{\lambda \phi_a}c_{12}^2  \ ,
        \nonumber \\
m_3 &=& \t m_3+ \frac{1}{2} \frac{D^2}{\lambda \phi_a}\ ,     \\
U_{e2}&=&s_{12}-  \frac{1}{2}\frac{D^2}{\lambda \phi_a}
\  \frac{c_{12}^2 s_{12}}{\t  m_2- \t m_1} \ , \nonumber\\
U_{\mu 3}&=& -\frac{1}{\sqrt{2}}\ + 
\frac{1}{2\sqrt{2}}\frac{D^2} {\lambda \phi_a} \left (\frac{c_{12}^2}
{ \t m_3- \t m_2} + \frac{s_{12}^2}
{ \t m_3- \t m_1}\right ), \nonumber\\
U_{e3}&=&
\frac{1}{2}\frac{D^2} {\lambda \phi_a} \left (\frac{c_{12}s_{12}}
{ \t m_3- \t m_2} - \frac{c_{12}s_{12}}
{ \t m_3- \t m_1}\right )\ . \nonumber
\label{solc}
\end{eqnarray}

 Taking same input parameters  in the case of $a=b=c=1$,
 we get numerical results as shown in Table 1.
In this case,  $m_3$,  $U_{\mu 3}$ and $U_{e3}$  also catch the effect of the
dark sector as well as   $m_1$, $m_2$ and  $U_{e2}$.
It is remarked that the non-vanishing  $U_{e3}$ is obtained
as well as the deviation from the maximal mixing of $U_{\mu 3}$.
The parameter $A$ and $k$ are  same one in the subsection 3.2.

%%%%%%%%%%%%%%%%%%%%%%%%%%%%%%%%%%%%%%%%
\section{Evolutions of Neutrino Masses and $w$}

If  $\phi^0_a$ is fixed, in other words, $D^2/\lambda$ is given,
we can  calculate the evolution of neutrino masses and $w$
by using eqs.(\ref{stationary2}) and (\ref{w}).
In order to see  these evolutions, we take
 $D^2/\lambda=1 \eV^2$  in eq.(\ref{dark}) in this section.

In the cases of $a=b=c=1$ and $a=1,\ b=c=0$, $m_3$ does not evolve 
because the dark sector does not contribute to the third family, 
while it does evolve a little in the case of $a=b=0,\ c=1$. 
Therefore, we show the evolution of neutrino mass $m_2$
 versus the redshift $z=T/T_0 -1$ for three cases in Figure 1.

As seen in Figure 1, the remarkable evolution of $m_2$
is found in the region of $z=0\sim 1$ for the case of $a=b=c=1$.
Evolutions of  cases of $a=1$, $b=c=0$
and  $a=b=0$, $c=1$ are  small compared with the one in the case of 
  $a=b=c=1$.
%%%%%%%%%%%%%%%%%%%%%%%%%%%%%%%%%%%%%%%%%%%%%%%%%%%%
%%%%%%%%%%%%%%%%%%%%%%%%%%%%
\begin{figure}
\begin{center}
\epsfxsize=10.  cm
\epsfbox{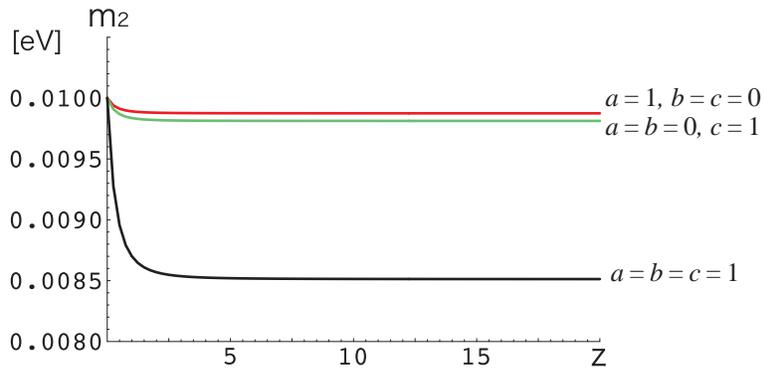}
\end{center}
\caption{Plot of $m_2$ versus the redshift $z$ for three cases.}
\end{figure}
%%%%%%%%%%%%%%%%%%%%%%%%%%%%
%%%%%%%%%%%%%%%%%%%%%%%%%%%%
\begin{figure}
\begin{center}
\epsfxsize=10.  cm
\epsfbox{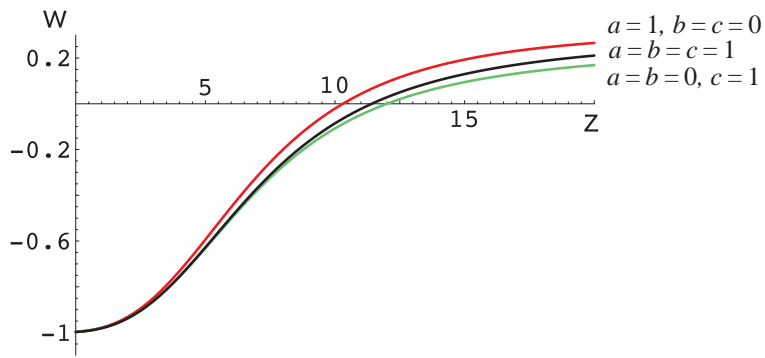}
\end{center}
\caption{Plot of the equation of state parameter $w$ versus
$z$ for  three cases.}
\end{figure}
%%%%%%%%%%%%%%%%%%%%%%%%%%%%
%%%%%%%%%%%%%%%%%%%%%%%%%%%%%%%%%%%%%%%%%%%%%%%%%%%%
Evolutions of $w$ versus $z$  are shown in  Figure 2.
These behaviors of $w$ are similar in three cases.
The  value of $w$ turns to positive around $z=10$,
which is somewhat different from the result  by Peccei \cite{Peccei},
in which the power-law potential was  also taken  
in  the one family model and then, $w$ becomes  positive  near $z=20$.

%%%%%%%%%%%%%%%%%%%%%%%%%%%%%%%%%%%%%%%%%%%%%%%%%%%%
%%%%%%%%%%%%%%%%%%%%%%%%%%%%%%%%%%%%%%%%%%%%%%%%%%%%
It was remarked \cite{stability} that the speed of sound, $c_s$ 
becomes imaginary in the non-relativistic limit at the present 
and then the Universe cease to accelerate.
 However, our prediction of $c_s^2$ is positive and  around  
 $0.1$ at the present epoch if the condition 
 $D^2/(\lambda \phi_a)\leq 5\times 10^{-4}$ is satisfied.
It means that  one  needs careful study  of the next-leading term
in  the non-relativistic limit of neutrinos.
%in the case of $m_1\sim {\cal O}(10^{-3}){\rm eV}$. 

In order to get the next-leading term in the non-relativistic limit,
 the function $ F(\xi _i)$ in eq.(\ref{Func}) is expanded  as
\begin{equation}
  F(\xi _i)=\frac{\xi_{\nu i}}{\pi ^2}
\left (\int _0^\infty\frac{dyy^2}{e^y+1}
  +\frac{1}{2 \xi_{\nu i}^2}\int _0^\infty\frac{dyy^4}{e^y+1} + \dots \right )
 \ .
\end{equation}
\noindent Then, the  speed of sound is given as 
\footnote{A paper including the  detail derivation of $c_s^2$  
  and its physical implication will appear  soon.}
\begin{eqnarray}
c_s^2&=&w +\frac{\dot w}{\dot\rho_{\mbox{dark}}}\ \rho_{\mbox{dark}} \\
&=&\frac{\displaystyle\sum_{i=1}^3
        \left(\frac{\partial m_{\nu i}}{\partial z}\hat n_\nu\right)}
        {\displaystyle\sum_{i=1}^3m_{\nu i}
        \frac{\partial \hat n_\nu}{\partial z}}
        +\frac{5}{3} f  \hat n_\nu \frac{
        \left[4T_0\left(\displaystyle\sum_{i=1}^3\frac{1}{\xi _i}\right)
        -T\left(\displaystyle\sum_{i=1}^3\frac{1}{\xi _i^2}
        \frac{\partial \xi _i}{\partial z}\right)\right]}
        {\displaystyle\sum_{i=1}^3m_{\nu i}
        \frac{\partial \hat n_\nu}{\partial z}}\ ,
\label{speed}
\end{eqnarray}
%\begin{eqnarray}
%c_s^2&=&w +\frac{\dot w}{\dot\rho_{\mbox{dark}}}\ \rho_{\mbox{dark}} \\
%&=&\frac{1}{\displaystyle\sum_{i=1}^3m_{\nu i}
%  \frac{\partial n_\nu ^{(0)}}{\partial z}}
%  \left ( {\displaystyle\sum_{i=1}^3
%    \left(\frac{\partial m_{\nu i}}{\partial z}n_\nu ^{(0)}\right)}
%      +\frac{5}{3} f  n_\nu ^{(0)}
%        {\left[4T_0\left(\displaystyle\sum_{i=1}^3\frac{1}{\xi _i}\right)
%        -T\left(\displaystyle\sum_{i=1}^3\frac{1}{\xi _i^2}
%        \frac{\partial \xi_i}{\partial z}\right)\right]}  \right )\ ,
%\label{speed}
%\end{eqnarray}
\noindent where
\begin{equation}
 \hat n_\nu  =\frac{T^3}{\pi ^2} \int _0^\infty\frac{dyy^2}{e^y+1}\ ,
\qquad  f =\frac{1}{2}
 \frac{\int _0^\infty\frac{dyy^4}{e^y+1}}
{\int _0^\infty\frac{dyy^2}{e^y+1}}\simeq 6.47 \ .
\end{equation}
\noindent In eq.(\ref{speed}), the time  derivative is replaced
 with $z$  derivative.
The first term in the right hand side of  eq.(\ref{speed})
is the leading term in the non-relativistic limit and is 
   negative definite because of 
${\partial m_{\nu i}}/{\partial z}<0$
and  ${\partial n_\nu ^{(0)}}/{\partial z}>0$.
 Since the numerator of the second term, which  is
  the next-leading term, is reduced to 
\begin{eqnarray}
 4T_0\left(\displaystyle\sum_{i=1}^3\frac{1}{\xi _i}\right)
        -T\left(\displaystyle\sum_{i=1}^3\frac{1}{\xi _i^2}
        \frac{\partial \xi _i}{\partial z}\right) 
= 5T_0\left(\displaystyle\sum_{i=1}^3\frac{1}{\xi _i}\right)
-\left(\displaystyle\sum_{i=1}^3\frac{1}{\xi _i^2}
        \frac{\partial m_{\nu i}}{\partial z}\right)  \ ,
\end{eqnarray}
the second term is  positive definite.
The magnitude of ${\partial m_{\nu i}}/{\partial z}$ is model dependent.
In our power-low potential for $V(\phi_a)$, 
the magnitude of the first term becomes very small. Then,  
 the second term is non-negligible one
in the case of $m_i\leq {\cal O}(10^{-2}){\rm eV}$.
If we take the condition $D^2/(\lambda \phi_a)\leq 5\times 10^{-4}$,
 the leading term is very small and then,
$c_s^2$ becomes positive due to the second term.
Actually, we checked numerically relative magnitude of
each term in the right hand side in eq.(\ref{speed}).
Thus,  the  positive $c_s^2$ is  understandable  in our model.

%%%%%%%%%%%%%%%%%%%%%%%%%%%%%%%%%%%%%%%%%%%%%%%%%%%%%%%%%%%%%%%%%%%%
%% &&  \sqrt{y^2+\xi _i^2}=\xi _i\sqrt{\left(\frac{y}{\xi _i}\right)^2+1}
%%%%%%%%%%%%%%%%%%%%%%%%%%%%%%%%%%%%%%%%%%%%%%%%%%%%%%%%%%%%%%%%%%%%
%%%%%%%%%%%%%%%%%%%%%%%%%%%%%%%%%%%%%%%%%%%%%%%%%%%%%%%%%%%%%%%%%%%%%%%%
%%%%%%%%%%%%%%%%%%%%%%%%%%%%%%%%%%%%%%%%%%%%%%%%%%%%%%%%%%%%%%%%%%%%%%%%
%%%%%%%%%%%%%%%%%%%%%%%%%%%%%%%%%%%%%%%%%%%%%%%%%%%%%%%%%%%%%%%%%%%%%%%%

%%%%%%%%%%%%%%%%%%%%%%%%%%%%%%%%%%%%%%%%%%%%%%%%%%%%%%%%%%%%%%%%%%%%%%%%
%%%%%%%%%%%%%%%%%%%%%%%%%%%%%%%%%%%%%%%%%%%%%%%%%%%%%%%%%%%
%%%%%%%%%%%%%%%%%%%%%%%%%%%%%%%%%%%%%%%%%%%%%%%%%%%%%%%%%%%
\section{Summary}

We have embedded the texture of the neutrino mass matrix 
with three families  
into the MaVaNs scenario. We take a typical texture of  active neutrinos,
 which is derived by the $D_4$ symmetry and predicts the maximal mixing 
of the atmospheric neutrino and the vanishing $U_{e3}$.
The effect of couplings among the dark sector 
and  active neutrinos are discussed by putting  the current cosmological data 
and the terrestrial neutrino experimental data.

It is found that  flavor  mixings evolve as well as  neutrino masses.
Especially,  $U_{e3}$ develops into the non-vanishing one
 and $\theta_{\rm atm}$ deviates from the maximal mixing
due to the dark sector in  the case of coupling $a=b=0,\ c=1$.
It is also remarked that 
the speed of sound, $c_s^2$ could be positive in our model.

Thus, 
the three family neutrino texture can be embedded into the MaVaNs scenario,
where the power-law potential of the  acceleron field, $\phi_a$ is taken.  
It is noticed that our results do not change so much even if
 we take the exponential  potential  of  $\phi_a$.
The related phenomena of our  flavor mixing varying  scenario
will be discussed  elsewhere.

%%%%%%%%%%%%%%%%%%%%%%%%%%%%%%%%%%%%%%%%%%%%%%%%%%%%%%%%%%%%%%%%%%%%%%%%%
 \section*{Acknowledgments}
  This work  is  supported by the Grant-in-Aid for Science Research,
 Ministry of Education, Science and Culture, Japan(No.16028205,17540243). 
%%%%%%%%%%%%%%%%%%%%%%%%%%%%%%%%%%%%%%%%%%%%%%%%%%%%%%%%%%%%%%%%%%%%%%%%%%

%%%%%%%%%%%%%%%%%%%%%%%%%%%%%%%%%%%%%%%%%%%%%%%%%%%%%
\end{document}